\pdfoutput=1

\documentclass{article}
\usepackage{ijcai19}

\usepackage{times}
\usepackage{soul}
\usepackage{url}
\usepackage[utf8]{inputenc}
\usepackage[small]{caption}
\usepackage{graphicx}
\usepackage{amsmath}
\usepackage{booktabs}
\usepackage{algorithm}
\usepackage{algorithmic}
\usepackage{multirow}
\usepackage{float}
\usepackage{wrapfig}
\usepackage{amssymb}

\title{Multi-Scale Quasi-RNN for Next Item Recommendation}

\author{
Chaoyue He$^{1,2}$\and
Yong Liu$^{1,2}$\and
Qingyu Guo$^{1,2}$\And
Chunyan Miao$^{1,2}$\\
\affiliations
 $^1$ School of Computer Science and Engineering, Nanyang Technological University\\
$^2$ LILY Research Centre, Nanyang Technological University\\
\emails
chaoyue001@e.ntu.edu.sg,
stephenliu@ntu.edu.sg,
qguo005@e.ntu.edu.sg,
ascymiao@ntu.edu.sg
}

\begin{document}

\maketitle

\begin{abstract}
How to better utilize sequential information has been extensively studied in the setting of recommender systems. To this end, architectural inductive biases such as Markov-Chains, Recurrent models, Convolutional networks and many others have demonstrated reasonable success on this task. This paper proposes a new neural architecture, multi-scale \textbf{Q}uasi-\textbf{R}NN for next item \textbf{Rec}ommendation (\textbf{QR-Rec}) task. Our model provides the best of both worlds by exploiting multi-scale convolutional features as the compositional gating functions of a recurrent cell. The model is implemented in a multi-scale fashion, i.e., convolutional filters of various widths are implemented to capture different union-level features of input sequences which influence the compositional encoder. The key idea aims to capture the recurrent relations between different kinds of local features, which has never been studied previously in the context of recommendation. Through extensive experiments, we demonstrate that our model achieves state-of-the-art performance on 15 well-established datasets, outperforming  strong competitors such as FPMC, Fossil and Caser absolutely by 0.57\%-7.16\% and relatively by 1.44\%-17.65\% in terms of MAP, Recall@10 and NDCG@10.

\end{abstract}

\graphicspath{ {figures/} }
\vspace{-5mm}
\section{Introduction}

In the era of information explosion, the filtering of relevant information forms the crux of a pleasant user experience. To this end, recommender systems are effective tools for solving this problem at hand. However, most of existing recommenders only utilize the history data as a set without considering the sequential features, so they can be limited to the scenarios where plenty of sequential information exists. Therefore, given the time and order sensitive nature of many domains (e.g. purchase or click logs, viewing history etc.), learning sequential item-based representations of users has become a widespread choice for many recommender systems \cite{quadrana2018sequence}. 

The search for suitable inductive biases for sequential representation learning has been a keen topic of interest. Markov Chain-based models have been a pillar in the field for long \cite{rendle2010factorizing,he2016fusing,he2017translation}, but their short-range and static representation problems in modeling user preferences are unavoidable. Recent state-of-the-art work in this domain have mainly revolved around neural models that rely on recurrence \cite{hidasi2015session}, convolution \cite{tang2018personalized} or many other architectures which offer different flavours and perspectives for learning sequential representations. To be more specific on recurrence and convolution, in \cite{hidasi2015session}, they used GRU-based RNN to capture user session preferences over items while in \cite{tang2018personalized}, they utilized two directions of convolutional filters to extract local features of item sequences. On one hand, recurrent models process the input chain of items in an auto-regressive fashion, in which each hidden state is conditioned on the previous while it ignores the union-level features among the sequence of items. On the other hand, convolutional models learn extractive, local features and inter-dependencies between adjacent items in a sequence while it ignores the recurrent relations among features. To balance the trade-offs, the idea of using convolution as gating for recurrence has been proposed. 

In fact, many hybrid CNN-RNN structures have been adopted in works across various domains including object recognition \cite{liang2015recurrent}, speech recognition \cite{sainath2015convolutional}, text modeling \cite{Wang:2017:HFT:3097983.3098140} etc. Whilst many gating mechanisms have been proposed like using single scale k-gram CNN \cite{bradbury2016quasi}, feed forward network \cite{lei2017training} or recurrent network \cite{Tay2018RecurrentlyCR} etc to generate gates. Therefore, this paper aims to combine the best of both worlds in a novel way, utilizing local, extractive, multi-scale convolution-based features as recurrent gating functions. To achieve so, we introduce a new neural architecture, multi-scale \textbf{Q}uasi-\textbf{R}NN for next item \textbf{Rec}ommendation (\textbf{QR-Rec}). The idea at hand is simple - gating functions are pre-learned via multi-scale convolution and then applied recursively in an auto-regressive fashion similar to a recurrent model. 

Therefore, in this paper, the contributions we made can be summarized as below:
\begin{enumerate}
  \item We proposed a novel architecture to extend Quasi-RNN with dynamic average pooling to a multi-scale fashion as the core building block. The advantage is that, it takes the user-item interaction sequence embeddings as input to the convolutional filters with 1 to full length of it. Therefore, different union-level features can be extracted and fed into a recurrent structure for better sequential representation learning. 
  \item We proposed a hierarchical summation aggregation (HSA) strategy to aggregate the outputs of each level in the whole architecture, which counts every unit cell and component output evenly contributed, retains more original information without scaling, and is shown to enhance the results to the maximum. 
  \item Experiments on 15 real-world datasets of Amazon Reviews are conducted and the results on 3 metrics outperform the state-of-the-art methods. We also investigated how the results were caused through several analyses of key model hyperparameters and different components.
\end{enumerate}

\vspace{-4mm}
\section{Related Work}
\vspace{-1mm}
In this section, we will review the work related to next item recommendation, hybrid CNN-RNN structures and various gating mechanisms for RNN.
\vspace{-1mm}
\subsection{Next Item Recommendation}
Next item recommendation is a sub field of sequential recommendation, which considers the order of single item with user general preference for the recommendation. These models are different from the general recommendation models which consider only the set of items without order, including the matrix factorization \cite{koren2009matrix} based models like TimeSVD++ \cite{koren2009collaborative}, PMF \cite{NIPS2007_3208} as well as the neural network based models like NCF \cite{he2017neural}, NeuRec \cite{Zhang_2018}, MPCN \cite{tay2018multi}, LRML \cite{tay2018latent}

Markov-Chain based models play as a role of pillar in this domain. In \cite{he2017translation}, TransRec mainly models third-order interactions between the user, the previously visited items and the next item to consume. FPMC \cite{rendle2010factorizing} integrates MF and first-order MC together while Fossil \cite{he2016fusing} integrates similarity-based methods with higher-order MCs as sequential recommenders.

In recent years, neural network based models become another pillar in recommender systems \cite{zhang2017deep}. Among those, GRU-based RNNs were proposed for the session-based recommendation, including GRU4Rec \cite{hidasi2015session} and GRU4Rec+ \cite{tan2016improved}, where GRU4Rec is the first proposed model for the problem while GRU4Rec+ improves the results by adding data augmentation and accounting for shifts in the input data distribution. In \cite{tang2018personalized}, CNN through two directions of filters was proposed as Caser and achieves the state-of-the-art for sequential recommendation task while in \cite{yuan2019simple}, NextItNet was proposed consisting of a stack of holed convolutional layers to efficiently increase the receptive fields. By integrating self-attention, SASRec \cite{kang2018self} uses it as main building block as well as captures long-term semantics to make predictions based on relatively few actions whilst AttRec \cite{zhang2019next} utilizes it to estimate relative weights of each item in user interaction trajectories for better learning user’s transient interests representations under metric learning framework. In \cite{chen2018sequential}, NARM was proposed to integrate an encoder with attention mechanism while in \cite{li2017neural}, they proposed RUM with memory network as the main building block to store and update users' historical records explicitly. In \cite{wu2018session}, graph neural network was utilized to model session data for session-based recommendation as SR-GNN. 
\vspace{-2mm}
\subsection{Hybrid CNN-RNN Structures} 
The neural architectures consisting of CNN and RNN actually take advantage of both worlds, where local features extraction and recurrence modeling can be fused together. In \cite{donahue2015long}, they proposed LR-CNs to integrate CNN extracting visual features followed by LSTM extracting sequence features on visual tasks involving sequences. In \cite{sainath2015convolutional}, they integrated CNNs, LSTMs and DNNs into a unified model as CLDNN with one followed by another on speech recognition tasks. In \cite{liang2015recurrent}, they proposed a recurrent CNN(RCNN) by incorporating recurrent connections into each convolution layer for object recognition. In \cite{Wang:2017:HFT:3097983.3098140}, conv-RNN was proposed to integrate Bi-RNN followed by convolutional layer as basic module for text modeling. In \cite{bradbury2016quasi}, they proposed Quais-RNN which uses CNN for generating gates for pooling on NLP tasks, which inspires this work.
\vspace{-1mm}
\subsection{RNN Gating Mechanisms} 
Beyond the vanilla RNN, LSTM and GRU are both commonly used variants with simple gating mechanisms to empower RNNs. In order to enhance RNNs, many research has also been done on gating mechanisms including Quais-RNN \cite{bradbury2016quasi} with single scale k-gram CNN to adaptively generate gates, SRU \cite{lei2017training} which uses a single-layer feed forward network with sigmoid activation function to generate gates and accelerates the training and RCRN \cite{Tay2018RecurrentlyCR} that learns the recurrent gating functions using recurrent networks etc. They were created for not only solving the issues of gradient explosion or vanishing for RNNs, but also improving accuracy and speed to convergence for the tasks involving sequences. 
\vspace{-3mm}
\section{Methodology}
\vspace{-1mm}
\begin{figure*}[!htbp]
\includegraphics[width=\textwidth]{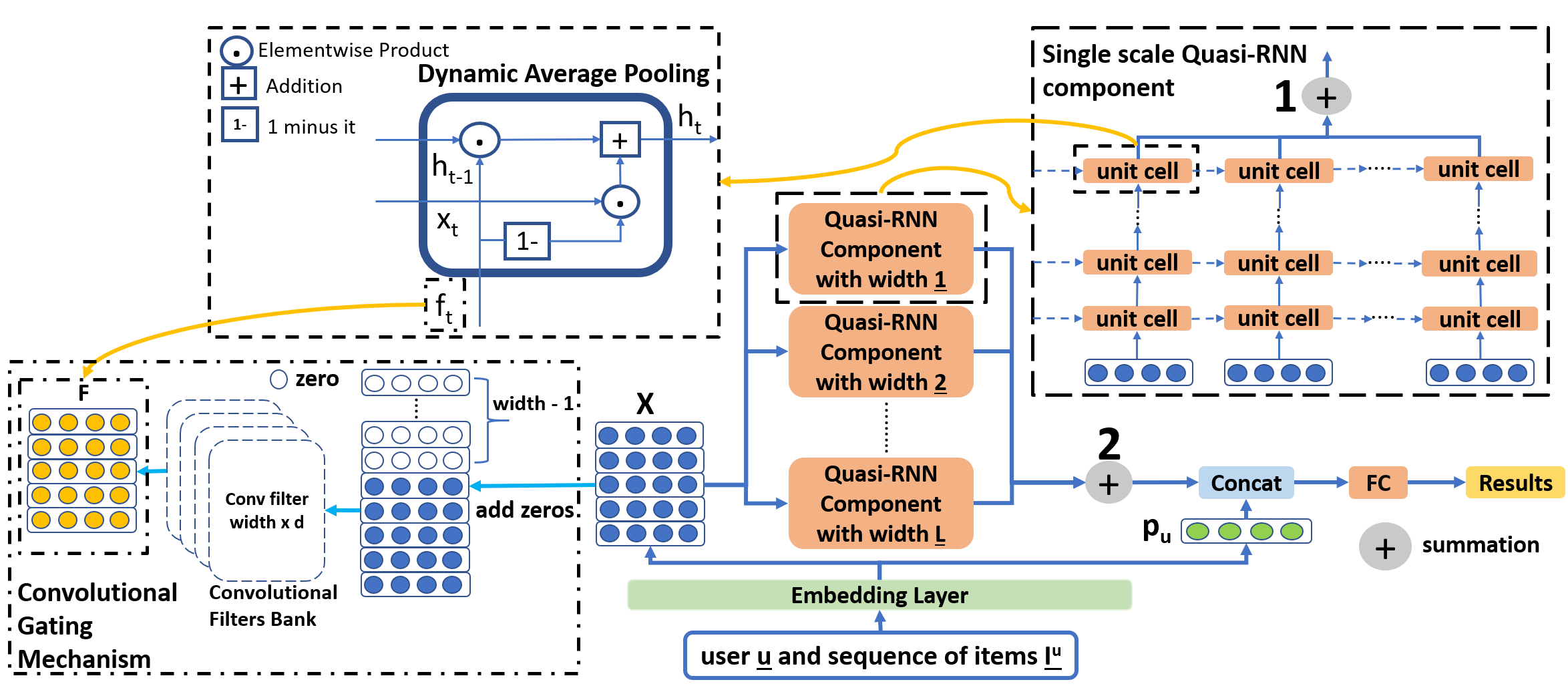}
\vspace{-7mm}
\caption{The whole architecture of the model}
\vspace{-6mm}
\end{figure*}
This next item recommendation problem is based on utilizing past user-item interaction sequences to make prediction for the most possible item he/she will interact with at next time step. Given $U$, the sets of users, $I$, the sets of items, each user is associated with a sequence of items he/she has interacted with in the past, $I^u$. By considering the user behaviour as a sequence of ordered items instead of a set of items, sequential recommendation shows its unique competitiveness over the general recommendation models. In this section, we will describe the model through an embedding layer, multi-scale Quasi-RNN and a prediction layer for final recommendation. 
\vspace{-1mm}
\subsection{Embedding Layer} 
\vspace{-1mm}
Through embedding layer, each item in the sequence of length $L$ will be mapped into a latent space with dimension $d$, as $x^u_{t-i+1} \in \mathbb{R}^d$ and $i \in [1,2,...,L]$ at time step $t-i+1$, the output of this layer $X^u_t \in \mathbb{R}^{d \times L}$ is a \textbf{sequence embedding matrix} of user $u$ before time step $t$ which concatenates all the item embeddings in the sequence.
\vspace{-1mm}
\begin{equation}
X^u_t = [x^u_{t-L+1};\cdots ;x^u_{t-1};x^u_{t}]
\vspace{-1mm}
\end{equation}
Also, each user $u$ is mapped into the same latent space to obtain a user profile representation as $p_u$ $\in$ $\mathbb{R}^{d}$.
\vspace{-2mm}
\subsection{Multi-Scale Quasi-RNN}
\vspace{-1mm}
As the core module of QR-Rec, through convolution on input sequences, each Quasi-RNN component with a specific filter width ($1$ to $L$) with latent dimension $d$, can then obtain different union-level features. We then introduce the subcomponents within the module one by one.
\vspace{-1mm}
\subsubsection{Convolution Gating Mechanism}
Convolutional gating mechanism aims to generate gates for the pooling layer. Like convolution layers in CNNs, it allows fully parallel computation across both mini batches and sequence dimension. Here we deduce the formulations below.
\vspace{-1mm}
\begin{equation}
F^u_t = \sigma(W^u_{f,t} \ast X^u_t) 
\vspace{-1mm}
\end{equation}
this is the simplest equation for width of 1. $W^u_{f,t} \in \mathbb{R}^{k \times d \times m}$ is the convolution filters bank, $m$ is the number of filters and it is set the same as $d$ for easy implementation. $\ast$ denotes a masked convolution along the timestep dimension. $X^u_t$ is the input sequence embedding matrix, $F^u_t \in \mathbb{R}^{k \times m} $ is the forget gate matrix and $\sigma$ is the sigmoid function. All the superscripts $u$ and subscripts $t$ indicate for user $u$ before time step $t$ here.
In fact, convolution filters of larger width can effectively compute higher union-level features at each timestep. So here we generalize the formulation for the gates for width of $N$.
\begin{equation}
\vspace{-3mm}
\small
f^{u,N}_t = \sigma(W_{f,t}^{u,1}x^u_{t-N+1}+...+W_{f,t}^{u,N-1}x^u_{t-1} + W_{f,t}^{u,N}x^u_{t})
\end{equation}
where $f^{u,N}_t \in \mathbb{R}^m$ is the forget gate vector, $W_{f,t}^{u,i} \in \mathbb{R}^{k \times d \times m}$ is the convolution filters bank where $i \in [1,2,...,N]$ for $u$ before time step $t$, and $x^u_i \in \mathbb{R}^d$ is the input vector where $i \in [t-N+1, t-N+2,...,t]$ indicates time step. 

\subsubsection{Dynamic Average Pooling}
We seek to obtain a function controlled by gates that can mix states across time steps, but also acts independently on each channel of the state vector. Here, we borrow the idea from dynamic average pooling \cite{pmlr-v48-balduzzi16} which only consists of a forget gate for controlling the information flow within the recurrent structure, and gives the general equation:

\vspace{-1mm}
\begin{equation}
\textbf{h}_t = \textbf{f}_t \odot \textbf{h}_{t-1} + (1 - \textbf{f}_t) \odot \textbf{x}_t
\end{equation}
where $\textbf{h}_t \in \mathbb{R}^d$ is the hidden state, $\textbf{f}_t \in \mathbb{R}^m$ is the forget gate and $\textbf{x}_t \in \mathbb{R}^d$ is the input item embedding at time step $t$.
Similar to pooling layers in CNNs, the pooling component here lacks trainable parameters and also allows fully parallel computation across mini batches and feature dimensions.

\subsubsection{Hierarchical Summation Aggregation}
Inside QR-Rec, aggregation needs placing at two points: 1) The output of a specific scale Quasi-RNN component; 2) The output of the whole multi-scale Quasi-RNN module. (The numbers \textbf{1} and \textbf{2} are shown on the Figure 1)
\vspace{-1mm}
\begin{equation}
\textbf{h}^u_{w,t} = \sum_{t_w=1}^Lh^u_{t_w}
\end{equation}
\vspace{-2mm}
\begin{equation}
\textbf{o}^u_t = \sum_{w=1}^L h^u_{w,t}
\vspace{-1mm}
\end{equation}
where $h^u_{t_w} \in \mathbb{R}^d$ is the hidden state obtained of unit cell of a Quasi-RNN component where $t_w \in [1,2,...,L]$ is time step for scale $w$,  $\textbf{h}^u_{w,t} \in \mathbb{R}^d $ is the output of a Quasi-RNN component with width $w$ where $w \in [1,2,...,L]$ for $u$ at $t$.
Then another summation aggregation for Quasi-RNN is performed to obtain $\textbf{o}^u_t \in \mathbb{R}^d$, the output for $u$ before $t$. By summation, all the features extracted from each scale component and every unit cell can be counted and information is retained to the maximum without scaling. 
\vspace{-1mm}
\subsection{Prediction Layer and Recommendation}
\vspace{-1mm}
$p_u$, the user embedding or profile, which represents user general preference, will be concatenated with $\textbf{o}^u_t$ to pass a fully-connected(\textbf{FC}) layer to obtain scores for candidate items. 
\vspace{-1mm}
\begin{equation}
y^u_t = W^u_{s,t}[o^u_t;p_u] + b^u_{s,t}
\vspace{-1mm}
\end{equation}
where $W^u_{s,t} \in \mathbb{R}^{2 \times d}$ and $b^u_{s,t} \in \mathbb{R}$ are parameters to learn for FC while $y^u_t$ is the final scores, all for $u$ before $t$.

Once upon the training is complete, we need $u$'s user embedding and his/her last $L$ item sequence embeddings as inputs. We will recommend top $N$ items with the highest scores of $y^u_t$. However, due to computation cost, we randomly sample \textbf{100} items from the negative items plus the item in the test set to form the candidate set for ranking computation.
\vspace{-3mm}
\section{Experiments}
\vspace{-1mm}
In order to show the effectiveness of QR-Rec, extensive experiments are performed followed by several qualitative analyses. The questions we aim to answer are as follows: 

\textbf{RQ1}- How does QR-Rec outperform other baselines? 

\textbf{RQ2}- How do key factors influence the performance? 

\textbf{RQ3}- What are different components' contributions?  
\vspace{-2mm}
\subsection{Datasets}
\vspace{-1mm}
\textbf{15} public datasets with high density from \textbf{Amazon Reviews\footnote{http://jmcauley.ucsd.edu/data/amazon/}} are used in the work as benchmark datasets. Table 1 shows the necessary statistics of the datasets. As for Amazon, it is a well-known giant online e-commerce website for selling products, and buyers may leave a review after they get the products. In our work, the review content is not necessary but it indicates the interaction of user-item pairs. We then perform preprocessing procedures below on the datasets : 
\begin{enumerate}
\vspace{-1.5mm}
\item Only capturing the interaction instances between users and items with rating no less than \textbf{3}, which is a reasonable threshold for judging preference.
\vspace{-1.5mm}
\item Determining the order of interactions in a user's historical sequence from timestep features. 
\vspace{-1.5mm}
\item Discarding user with less than \textbf{10} interactions to keep sequences for each user  reasonably and comparably long. 
\end{enumerate}
\begin{table}[!htbp]
\vspace{-5mm}
  \caption{Statistics of the datasets}
  \vspace{-3mm}
  \label{tab:Statistics}
  \small
  \centering
  \resizebox{\columnwidth}{!}{%
  \begin{tabular}{rrrrrrr}
    \toprule
    \textbf{Dataset}  & \textbf{\#users} & \textbf{\#items} & \textbf{\#interactions} & \textbf{sparsity} \\
    \midrule
    Electronics & 1530 & 29381 & 101436 & 99.77\% \\
    Tools / Home & 2908 & 9565 & 45786 & 99.84\%\\
    Cell Phones  & 2440 & 8152 & 36246 & 99.82\%\\
    Beauty & 4204 & 11164 & 75103 & 99.84\%\\
    Health / Care & 7398 & 17257 & 138083 & 99.89\%\\
    Video Games & 5048 & 10344 & 96344 & 99.82\%\\
    Toys / Games & 3723 & 11259 & 65877 & 99.84\%\\
    Sports / Outdoors & 6915 & 17337 & 109581 & 99.91\% \\
    Pet Supplies & 3096 & 7854 & 46418 & 99.81\% \\
    Digital Music & 1551 & 3510 & 35991 & 99.34\%\\
    Grocery / Food & 3786 & 8424 & 74656 & 99.77\% \\
    Instant Video & 540 & 1527 & 8133 & 99.01\%\\
    Home / Kitchen & 2247 & 19181 & 72781 & 99.83\% \\
    Android Apps & 2541 & 10103 & 81749 & 99.68\% \\
    Baby & 3488 & 6617 & 53144 & 99.77\% \\ 
  \bottomrule
  \end{tabular}}
  \vspace{-5mm}
\end{table}
\vspace{-1mm}
\subsection{Compared Baselines}
\begin{table*}[!ht]
\small
  \caption{Experiment Results}
  \vspace{-2mm}
  \label{tab:Results}
  \small

  \resizebox{\textwidth}{!}{%
  \begin{tabular}{r|r|rrrrrrrr|r|r}
    \toprule
    \textbf{Dataset}  & \textbf{Metric} & \textbf{PopRec} & \textbf{BprMF} &\textbf{FMC} &\textbf{FPMC} & \textbf{Fossil} & \textbf{GRU4Rec} & \textbf{Caser} & \textbf{QR-Rec} & \textbf{Abs Gain} &  \textbf{Rel Gain} \\
    \hline
    \multirow{3}{*}{Electronics} & MAP & 0.1739 & 0.3086 & 0.3471 & \underline{0.4349} & 0.4287 & 0.2253 & 0.4307                              & \textbf{0.4846} & 4.97\% &11.43\% \\
                                 & Recall@10 & 0.3562 & 0.6046 & 0.5595 & \underline{0.6739} & 0.6693 & 0.5150 & 0.6647 & \textbf{0.7301} & 5.62\% & 8.34\% \\
                                 & NDCG@10 & 0.2359 & 0.4332 & 0.4473 & 0.5488 & \underline{0.5492} & 0.3208 & 0.5468 & \textbf{0.6105} & 6.13\% & 11.16\% \\
    \hline
    \multirow{3}{*}{Tools / Home} & MAP & 0.1321 & 0.2292 & 0.1576 & \underline{0.2515} & 0.2254 & 0.1580 & 0.1227                              & \textbf{0.2804} & 2.89\% & 11.49\% \\
                                 & Recall@10 & 0.2634 & 0.4041 & 0.2999 & 0.3948 & \underline{0.4099} & 0.2909 & 0.3999 & \textbf{0.4570} & 4.71\% & 11.49\% \\
                                 & NDCG@10 & 0.1717 & 0.2986 & 0.1954 & \underline{0.3032} & 0.2940 & 0.1994 & 0.2951 & \textbf{0.3468} & 4.36\% & 14.38\%\\
    \hline
    \multirow{3}{*}{Cell Phones} & MAP & 0.1456 & 0.3312 & 0.2193 & 0.3378 & 0.3207 & 0.1969 & \underline{0.3400} &                             \textbf{0.4000} & 6.00\% & 17.65\%\\
                                 & Recall@10 & 0.3045 & 0.5369 & 0.4291 & 0.5291 & \underline{0.5721} & 0.3902 & 0.5467 & \textbf{0.6197} & 4.76\% & 8.32\% \\
                                 & NDCG@10 & 0.1956 & 0.4182 & 0.2852 & 0.4157 & \underline{0.4284} & 0.2603 & 0.4239 & \textbf{0.5000} & 7.16\% & 16.71\%\\
    \hline
    \multirow{3}{*}{Beauty} & MAP & 0.1309 & 0.3178 & 0.3372 & \underline{0.3828} & 0.3512 & 0.2661 & 0.3742 &                                \textbf{0.4025} & 1.97\% & 5.15\% \\
                                 & Recall@10 & 0.3390 & 0.5618 & 0.4900 & \underline{0.5657} & 0.5566 & 0.4715 & 0.5614 & \textbf{0.5978} &3.21\% &5.67\%\\
                                 & NDCG@10 & 0.1934 & 0.4176 & 0.4032 & \underline{0.4669} & 0.4471 & 0.3479 & 0.4605 & \textbf{0.4925} & 2.56\% & 5.48\%\\
    \hline
    \multirow{3}{*}{Health / Care} & MAP & 0.1604 & 0.2963 & 0.3002 & \underline{0.3638} & 0.3473 & 0.2916 &                                   0.3345 & \textbf{0.3770} & 1.32\% & 3.63\% \\
                                 & Recall@10 & 0.3713 & 0.5196 & 0.4840 & 0.5304 & \underline{0.5485} & 0.4849 & 0.5180 & \textbf{0.5741} & 2.56\% & 4.67\% \\
                                 & NDCG@10 & 0.2193 & 0.3867 & 0.3740 & \underline{0.4380} & 0.4361 & 0.3660 & 0.4155 & \textbf{0.4662} & 2.81\% & 6.41\% \\
    \hline
    \multirow{3}{*}{Video Games} & MAP & 0.1804 & 0.3832 & 0.3213 & \underline{0.4027} & 0.3909 & 0.2850 & 0.3938 &                              \textbf{0.4278} & 2.51\% & 6.23\% \\
                                 & Recall@10 & 0.3794 & \underline{0.6680} & 0.5588 & 0.6664 & 0.6664 & 0.5450 & 0.6638 & \textbf{0.6949} & 2.69\% & 4.03\% \\
                                 & NDCG@10 & 0.2447 & 0.5052 & 0.4116 & \underline{0.5128} & 0.5091 & 0.3820 & 0.5081 & \textbf{0.5445} & 3.17\% & 6.18\%\\
    \hline
    \multirow{3}{*}{Toys / Games} & MAP & 0.1163 & 0.2796 & 0.2599 & \underline{0.3544} & 0.2855 & 0.1968 & 0.3204                              & \textbf{0.3641} & 0.97\% & 2.74\%\\
                                 & Recall@10 & 0.2834 & 0.5004 & 0.4279 & 0.5195 & \underline{0.5428} & 0.4021 & 0.5205 & \textbf{0.5713} & 2.85\% & 5.25\% \\
                                 & NDCG@10 & 0.1600 & 0.3661 & 0.3259 & \underline{0.4255} & 0.3882 & 0.2661 & 0.4070 & \textbf{0.4552} & 2.97\% & 6.98\%\\
    \hline
    \multirow{3}{*}{Sports / Outdoors} & MAP & 0.1570 & 0.2770 & 0.2073 & \underline{0.2777} & 0.2475 & 0.1916 & 0.2516 & \textbf{0.2872} & 0.95\% & 3.42\% \\
                                 & Recall@10 & 0.3145 & 0.4782 & 0.3564 & \underline{0.4797} & 0.4525 & 0.3748 & 0.4479 & \textbf{0.4943} & 1.46\% & 3.04\% \\
                                 & NDCG@10 & 0.2057 & \underline{0.3545} & 0.2634 & 0.3542 & 0.3235 & 0.2502 & 0.3230 &\textbf{0.3674} & 1.29\% & 3.64\% \\
    \hline
    \multirow{3}{*}{Pet Supplies} & MAP & 0.1539 & \underline{0.2843} & 0.2213 & 0.2819 & 0.2529 & 0.1956 & 0.2650                              & \textbf{0.3149} & 3.06\% & 10.76\% \\
                                 & Recall@10 & 0.3272 & \underline{0.5081} & 0.3947 & 0.4916 & 0.4564 & 0.3656 & 0.4616 & \textbf{0.5233} & 1.52\% & 2.99\%\\
                                 & NDCG@10 & 0.2119 & \underline{0.3747} & 0.2815 & 0.3644 & 0.3347 & 0.2544 & 0.3422 & \textbf{0.3980} & 2.33\% & 6.22\% \\
    \hline
    \multirow{3}{*}{Digital Music} & MAP & 0.1237 & 0.2998 & 0.2358 & 0.3126 & 0.3169 & 0.2184 &                                               \underline{0.3227} & \textbf{0.3683} & 4.56\% & 14.13\% \\
                                 & Recall@10 & 0.2805 & \underline{0.6009} & 0.4429 & 0.5996 & 0.6164 & 0.4513 & 0.5893 & \textbf{0.6254} & 2.45\% & 4.08\% \\
                                 & NDCG@10 & 0.1661 & 0.4238 & 0.3061 & 0.4208 & \underline{0.4326} & 0.2982 & 0.4275 & \textbf{0.4757} & 4.31\% & 9.96\% \\
    \hline
    \multirow{3}{*}{Grocery / Food} & MAP & 0.1763 & 0.2542 & 0.3424 & \underline{0.3961} & 0.3026 & 0.2733 & 0.3773                              & \textbf{0.4018} & 0.57\% & 1.44\% \\
                                 & Recall@10 & 0.4762 & 0.5742 & 0.5148 & 0.5734 & \underline{0.5763} & 0.4997 & 0.5647 & \textbf{0.6157} & 3.94\% & 6.84\% \\
                                 & NDCG@10 & 0.2711 & 0.3744 & 0.4210 & \underline{0.4796} & 0.4191 & 0.3692 & 0.4651 & \textbf{0.5027} & 2.31\% & 4.82\% \\
    \hline
    \multirow{3}{*}{Instant Video} & MAP & 0.1421 & 0.2212 & 0.1963 & 0.2414 & 0.2107 & 0.1596 &                                               \underline{0.2519} & \textbf{0.2825} & 3.06\% & 12.15\% \\
                                 & Recall@10 & 0.2870 & 0.4463 & 0.3722 & 0.4519 & 0.4556 & 0.3426 & \underline{0.4852} & \textbf{0.5519} & 6.67\% & 13.75\% \\
                                 & NDCG@10 & 0.1877 & 0.3085 & 0.2544 & 0.3151 & 0.2937 & 0.2169 & \underline{0.3387} & \textbf{0.3808} & 4.21\% & 12.43\% \\
    \hline
    \multirow{3}{*}{Home / Kitchen} & MAP & 0.1565 & 0.3093 & 0.3006 & \underline{0.4028} & 0.3448 & 0.2160 &                                   0.3597 & \textbf{0.4148} & 1.20\% & 2.98\% \\
                                 & Recall@10 & 0.3507 & 0.5741 & 0.4998 & \underline{0.5937} & 0.5906 & 0.4117 & 0.5536 & \textbf{0.6075} & 1.38\% & 2.32\%\\
                                 & NDCG@10 & 0.2160 & 0.4216 & 0.3831 & \underline{0.4839} & 0.4560 & 0.2897 & 0.4540 & \textbf{0.5107} & 2.68\% & 5.54\% \\
    \hline
    \multirow{3}{*}{Android Apps} & MAP & 0.2002 & 0.3330 & 0.2702 & \underline{0.3670} & 0.3479 & 0.2247 & 0.3420 &                              \textbf{0.4022} & 3.52\% & 9.59\% \\
                                 & Recall@10 & 0.4099 & 0.6686 & 0.5455 & \underline{0.6690} & 0.6623 & 0.4526 & 0.6218 & \textbf{0.6808} & 1.18\% & 1.76\% \\
                                 & NDCG@10 & 0.2648 & 0.4682 & 0.3660 & \underline{0.4972} & 0.4820 & 0.2998 & 0.4542 & \textbf{0.5212} & 2.40\% & 4.83\%\\
    \hline
    \multirow{3}{*}{Baby} & MAP & 0.1458 & 0.1979 & 0.1797 & \underline{0.2121} & 0.1955 & 0.1708 & 0.1963 &                                   \textbf{0.2322} & 2.01\% & 9.48\% \\
                                 & Recall@10 & 0.3191 & 0.4120 & 0.3458 & \underline{0.4131} & 0.4074 & 0.3632 & 0.3698 & \textbf{0.4384} & 2.53\% & 6.12\%\\
                                 & NDCG@10 & 0.1958 & 0.2761 & 0.2302 & \underline{0.2831} & 0.2717 & 0.2321 & 0.2556 & \textbf{0.3108} & 2.77\% & 9.78\% \\
                                 
    
  \bottomrule
  \end{tabular}}
  \vspace{-4mm}
\end{table*}
QR-Rec is compared with 7 following baselines.

\textbf{PopRec} recommends based on the popularity(frequency) of items across the whole dataset. \textbf{BprMF} uses MF under Bayesian personalized ranking framework \cite{rendle2009bpr} for non-sequential recommenders. \textbf{FMC} does MF on first-order MC and recommends based on the score. \textbf{FPMC} \cite{rendle2010factorizing} adds user general preference to FMC and treats each basket of only one item in this work. \textbf{Fossil} \cite{he2016fusing} uses similarity based model instead of LFM for modeling user general preferences based on high-order MCs. \textbf{GRU4Rec} \cite{hidasi2015session} uses GRUs to model user action sequences for session-based recommendation. We regard users' interaction sequences as sessions. Here, we don't use GRU4Rec+ since it adopts a different loss function and sampling strategy from all the others in the paper. \textbf{Caser} \cite{tang2018personalized} utilizes two directions of CNNs to extract union-level and individual-level features on the sequence embedding matrix, and achieves the state-of-the-art sequential recommendation performance.
\vspace{-1mm}
\subsection{Experimental Setup}
This section is composed of model training, evaluation protocol and implementation detail subsections as below.
\vspace{-2mm}
\subsubsection{Model Training}
We select \textbf{binary cross-entropy} loss as the objective function to minimize:
\vspace{-1mm}
\begin{equation}
\displaystyle l = \sum_{u}(\sum_{i}-\log(\sigma(y_{i,t}^u))+\sum_{j}-log(1-\sigma(y_{j,t}^u)))
\vspace{-3mm}
\end{equation} 
where $\sigma$ is sigmoid function, $i$ is target item while $j$ is negative item. We randomly sample \textbf{3} negative items for each target item during training, and this follows the works \cite{he2016fusing,rendle2010factorizing,tang2018personalized}.

The model parameters, {Item and User Embeddings, Convoluational Weights and other neural network parameters} are all learned by minimizing the objective function on the training set without pre-training, while the hyperparameters {Dropout Ratio, Learning Rate, L2 Regularization, Number of Layers} are tuned on the validation set via grid search. The experiments are mainly conducted on 4 Nvidia Tesla 80K GPUs and partly on a single RTX 2080Ti GPU. 
\vspace{-2mm}
\subsubsection{Evaluation Protocol} \textbf{Leave-One-Out}: For validation, we use each user's first item until third last for training and the second last for testing. For testing, we use each user's first item until second last for training and the last item for testing.

\textbf{Evaluation Metrics}: \textbf{MAP} (Mean Average Precision), \textbf{Recall@10}, \textbf{NDCG@10} are selected. We train the model for \textbf{20} epochs at first and continue the training if there are improvements on any metrics. The evaluation scores are recorded every epoch and final reporting score is based on the best validation setting's corresponding testing case.
\vspace{-1mm}
\subsubsection{Implementation Details}

We implement the models and whole experiment using PyTorch. The optimizer we use for all the models is Adam \cite{kingma2014adam}, which is an variant of Stochastic Gradient Descent (SGD) with learning rate tuned through \textbf{[0.001, 0.0003, 0.0001]} and finally fixed at \textbf{0.001}. The batch size is fixed at \textbf{512} and the L2 regularization is fixed at \textbf{1e-6} across all the models. We add Dropout \cite{srivastava2014dropout} to the models and it is tuned at \textbf{\{0.2,0.5\}}. The number of layers is tuned through \textbf{[1,2,3,4]}. After the exploration of different latent dimensions \textbf{[16,32,64,128,256]} and sequence length \textbf{[4,5,6,7,8]} in the analyses, we set the embedding dimension fixed at \textbf{128} and sequence length of \textbf{5} for fair comparison. All model parameters are initialized from normal distribution.  
\vspace{-2mm}
\subsection{Experiment Results}
\vspace{-1mm}
Table 2 recorded the results of 7 baselines and QR-Rec. The best performer of each row is highlighted in boldface while the second best with underline. The second last column \textbf{Abs Gain} and the last column \textbf{Rel Gain} are calculated as \textit{QR-Rec - second best performer} and $\frac{\textit{Abs Gain}}{second\:best\:performer}$, and they are the absolute and relative gain respectively. QR-Rec is the best performer on all datasets with all the metrics. The absolute gain on all datasets can be from 0.57\% to 7.16\% while the relative gain can be 1.44\% to 17.65\%, since the datasets we use are all very dense ( $>$ \textbf{99\%}), so we can confirm the power of QR-Rec on dense datasets. Therefore, it reveals the recurrent structure of extractive local union-level features do exist there and it can be handled very well with QR-Rec.

Among all, although Caser is one of the state-of-the-arts using CNN, FPMC and Fossil as MC-based sequential recommenders are still more powerful on many datasets, and they also outperform over non-sequential recommender like BprMF. Also, by including user profile, FPMC and Fossil outperform over FMC, which shows that personalization can truly enhance the performance. As for GRU4Rec, which is more used in session-based recommendation without personalization, it underperforms below many baselines.
\vspace{-3mm}
\section{Model Analysis and Discussion}
\vspace{-1mm}
We conduct several analyses to show more intuitions about QR-Rec here. \textbf{NDCG@10} is selected for all analyses, because it considers both prediction accuracy and ranking.
\vspace{-2mm}
\subsection{Key Hyperparameter Analysis}
We select 2 datasets for \textbf{sequence length} and \textbf{latent dimension} analyses, and record the results in Figure 2.
\begin{figure}[!htbp]
\vspace{-4mm}
\includegraphics[width = \columnwidth]{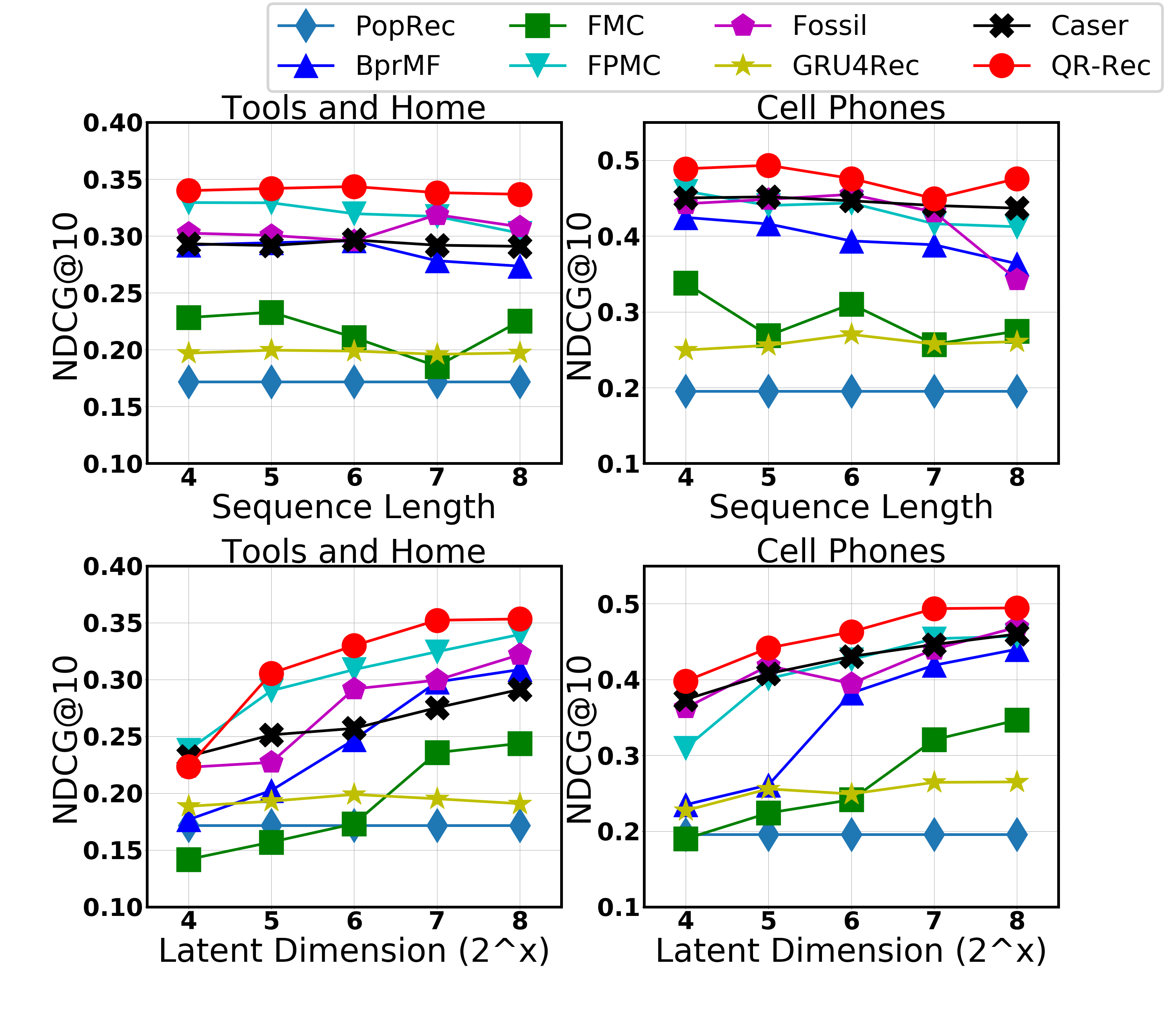}
\vspace{-10mm}
\caption{Key Hyperparameter Analysis}
\vspace{-6.5mm}
\end{figure}
\subsubsection{Sequence Length}
By assumption, only recent items can influence user's next item most significantly, so we choose a set of comparably short sequence lengths   [4,5,6,7,8] for analysis. In the first row in Figure 2, we show the performance of QR-Rec is always the best and stabilizes across different lengths. Thus, selecting 5 will be enough to represent general cases.
\vspace{-1mm}
\subsubsection{Latent Dimension}
Latent dimension is a key factor to determine the evaluation results as well as the computation costs. Here we tested on the latent dimension of [16,32,64,128,256]. In the second row in Figure 2, we show the performance for QR-Rec is almost always the best. Furthermore, all models almost increase from low to high dimension monotonically, but the difference between 128 and 256 are not obvious, so we choose 128 for the whole experiments as an optimal selection for balancing performance and costs.
\vspace{-3mm}
\subsection{Output Gate Analysis}

We compare our dynamic average pooling consisting of only a \textbf{forget gate} with a more complicated setting consisting of one more \textbf{output gate}, which is inspired by the original Quais-RNN structure. 
\vspace{-2mm}
\begin{equation}
O^u_t = \sigma(W^u_{o,t} \ast X^u_t)
\vspace{-7mm}
\end{equation}

\begin{equation}
\small
o^{u,N}_t = \sigma(W_{o,t}^{u,1}x^u_{t-N+1} + ... + W_{o,t}^{u,N-1}x^u_{t-1} + W_{o,t}^{u,N}x^u_{t})
\vspace{-5mm}
\end{equation}

\begin{equation}
\textbf{c}_t = \textbf{f}_t \odot \textbf{c}_{t-1} + (1 - \textbf{f}_t) \odot \textbf{x}_t
\vspace{-1mm}
\end{equation}
\begin{equation}
\textbf{h}_t = \textbf{o}_t \odot \textbf{c}_t
\vspace{-1mm}
\end{equation}
The way to obtain an output gate is similar to the forget gate. Then for the pooling part, we need to element-wise multiply the output gate to the hidden state. We show the comparison on 4 datasets in Table 3, and find that our simpler version is better. The reason behind should be that, since our setting of sequence length is short, the more complicated pooling can be redundant so it fails to beat dynamic average pooling.
\vspace{-1mm}
\begin{table}[!htbp]
\vspace{-3mm}
  \caption{Output Gate Analysis}
  \vspace{-2mm}
  \label{tab:output gate analysis}
  \small
  \centering
  \resizebox{\columnwidth}{!}{%
  \begin{tabular}{r|rrrr}
    \toprule
      Datasets & Tools / Home & Beauty & Digital Music & Pet Supplies \\
    \midrule
    with $O$ & 0.3424 & 0.4892 & 0.4514 & 0.3957 \\
    w/o $O$ & \textbf{0.3468} & \textbf{0.4925} & \textbf{0.4757} &  \textbf{0.3980} \\
  \bottomrule
  \end{tabular}}
  \vspace{-5mm}
\end{table}

\subsection{Aggregation Strategy Analysis}
 We can use the \textbf{last hidden state} of each scale Quasi-RNN component (\textbf{L}) at 1) and taking \textbf{summation operation} (\textbf{S}) or taking \textbf{mean operation} (\textbf{M}) at 1) and 2). We record the results of all possible combinations in Table 4,
\begin{table}[!htbp]
\vspace{-3mm}
  \caption{Aggregation Strategy Analysis}
  \vspace{-2mm}
  \label{tab:aggregation strategy analysis}
  \small
  \centering
  \resizebox{\columnwidth}{!}{%
  \begin{tabular}{r|rrrr}
    \toprule
      Datasets & Tools / Home & Beauty & Digital Music & Pet Supplies \\
    \midrule
    L+S  & 0.2830 & 0.4513 & 0.4082 & 0.3491\\
    L+M  & 0.2692 & 0.4328 & 0.4397 & 0.3317\\
    S+M(M+S) & 0.3463 & 0.4923 & 0.4558 & 0.3757\\
    M+M & 0.2587 & 0.4244 & 0.4238 & 0.3434 \\
    HSA(S+S) & \textbf{0.3468} & \textbf{0.4925} & \textbf{0.4757} & \textbf{0.3980} \\
  \bottomrule
  \end{tabular}}
  \vspace{-4.5mm}
\end{table}
 We find through two summation aggregations, named as Hierarchical Summation Aggregation (\textbf{HSA}), the results are the best. The reason behind is probably due to the summation can consider all the features extracted from each scale component and every unit cell, then there is no useful information loss.
 \vspace{-2mm}
\subsection{User Profile Analysis}
In Table 5, we show that on 4 datasets, by integrating $p_u$, the results can be enhanced, and we also show the model without $p_u$ and how $p_u$ factor only contributes to the results. By comparison, we can conclude that our core module improves the results to a large margin. 

\begin{table}[htb]
\vspace{-3mm}
  \caption{User Profile Analysis}
  \vspace{-2mm}
  \label{tab:user profile analysis}
  \small
  \centering
  \resizebox{\columnwidth}{!}{%
  \begin{tabular}{r|rrrr}
    \toprule
      Datasets & Tools / Home  & Beauty & Digital Music & Pet Supplies \\
    \midrule
    $p_u$ only & 0.2923 & 0.4002 & 0.4390 & 0.3678 \\
    QR-Rec w/o $p_u$ & 0.3382 & 0.4879 & 0.4450 & 0.3570 \\
    QR-Rec & \textbf{0.3468} & \textbf{0.4925} & \textbf{0.4757} & \textbf{0.3980}\\
  \bottomrule
  \end{tabular}}
  \vspace{-6mm}
\end{table}
\subsection{Scale Analysis}
In Table 6, we record the results on 4 datasets to show the effectiveness of our multi-scale setting. By comparing QR-Rec with single scale Quasi-RNN of scale from 1 to 5 under the default setting, we see clearly the superior performance QR-Rec can achieve over single scale ones, and it reconfirms the power of multi-scale setting.
\begin{table}[htb]
\vspace{-3mm}
  \caption{Scale Analysis}
  \vspace{-2mm}
  \label{tab:scale analysis}
  \small
  \centering
  \resizebox{\columnwidth}{!}{%
  \begin{tabular}{r|rrrr}
    \toprule
      Datasets & Tools / Home  & Beauty & Digital Music & Pet Supplies \\
    \midrule
    Quasi-RNN(w=1) & 0.2923 & 0.4398 & 0.4400 & 0.3518 \\
    Quasi-RNN(w=2) & 0.2659 & 0.4344 & 0.4435 & 0.3389 \\
    Quasi-RNN(w=3) & 0.2649 & 0.4267 & 0.4344 & 0.3323 \\
    Quasi-RNN(w=4) & 0.2644 & 0.4319 & 0.4318 & 0.3330 \\
    Quasi-RNN(w=5) & 0.2707 & 0.4320 & 0.4346 & 0.3254 \\
    QR-Rec  & \textbf{0.3468} & \textbf{0.4925} & \textbf{0.4757} & \textbf{0.3980}\\
  \bottomrule
  \end{tabular}}
  \vspace{-4mm}
\end{table}
\vspace{-3mm}
\section{Conclusion}
\vspace{-1mm}
In the paper, we proposed QR-Rec to combine the power of convolution and recurrence worlds for next item recommendation task. Through integration of multi-scale Quasi-RNN module, the recurrent features of different union-level features of user-item sequences can be captured more suitably. Extensive experiments show that QR-Rec is strong across various datasets and several quantitative analyses reveal the intuitions behind model components and hyperparameters.

\bibliographystyle{named}
\bibliography{ijcai19}
\end{document}